%% file: chapter.tex
\ifx\weAreInMain\undefined                                                      %
\documentclass[a4paper]{book}                                                   %
\input{main/macros.tex}                                                   %
\input{./macros-chapter.tex}                                                   
\input{./style-chapter.tex}                                             %
\usepackage{hyperref}
\begin{document}                                                         %
  \renewcommand{\chapterroot}{./}                                         %
\fi                                                                             %

%

\chapter*{Cell behavior in the face of uncertainty}
\label{UNC}

\includechapterauthors{David Lacoste, Olivier Rivoire, and David S. Tourigny}

\begin{chapterhighlights}

\begin{itemize}
\item Organisms that grow and survive in uncertain environments may need to change their physiological state as the environment changes.
\item When the environment is uncertain, one strategy known as bet-hedging is to make these changes randomly and independently of the environment, to ensure that at least part of the population is well adapted.
\item Organisms that collect information from their environment may also use this information to modulate their changes of physiological states.
\item We review these different strategies and point out parallels with the theory of optimal financial investments.
\end{itemize}

\end{chapterhighlights}

To a large extent, the content of the textbook \href{https://principlescellphysiology.org/book-economic-principles/index.html}{Economic Principles in Cell Biology} prior to the current chapter has dealt with models of microorganisms under the implicit assumption that the dynamics of both environmental factors and intracellular components are deterministic, and that behavior is optimized uniformly across cells in a population. On longer time scales however, natural selection also acts on populations and these populations may encounter environments that fluctuate across both time and space.
Under these conditions, natural selection may not favor a homogeneous deterministic cellular response across the population, but rather select for a certain level of population diversity and heterogeneity, including behaviors arising from mechanisms that are fundamentally stochastic. Stochasticity is inherent to intracellular processes such as gene expression and signal transduction due to the small number of molecules that they involve. It is often referred to as ``noise'', but this terminology can be misleading because stochasticity may also fulfill an essential role in cellular function and survival, for example during growth in uncertain environmental conditions. The purpose of this chapter is to highlight this role, introduce the mathematical models necessary for understanding it, and draw a new economic analogy with problems of investment in finance.

Before expanding upon the role that uncertainty plays in shaping cellular behavior, we briefly point out some general limitations of deterministic models based on optimal regulation of behavior in time as described in the chapter {\em Optimal cell behavior in time}. In that chapter, it was assumed that microorganisms have evolved, under selective pressures exerted by the environment, to optimize a specific objective criterion or combination of objective criteria that were shared by all cells of a population. This assumption was then incorporated into an optimal control framework to explain how cellular behavior (e.g., enzyme expression) is optimally regulated in time depending on deterministic interactions between a microbial population and its environment. 
In particular, we consider optimal control strategies across a prescribed time window.
Defining optimality in such case assumes the organism has perfect information on how the environment will change (including in response to actions taken) over time. In an uncertain environment, this information is simply not available. An alternative is instantaneous optimization of growth rate at each time point but this is a shortsighted strategy that excludes any partial information on future environmental states that the organism may have acquired over the course of evolution. Such deterministic models may be suitable for deterministically changing environments, but cannot account for stochastic behaviors that may be advantageous to population growth in uncertain environments.                     

In this chapter, it will be shown how principles of optimality can be formulated to study the behavior of organisms growing under uncertainty. Unlike the deterministic setting however, optimality will instead need to be defined in terms of probabilities and expected returns. Analogous to the general unification of deterministic models for cellular behavior using an optimal control theory framework, models including uncertainty are unified by the subject of {\em stochastic optimal control}. Beyond biology, this subject has wide-reaching applications to engineering but the most relevant analogy is with finance where stochastic strategies of portfolios diversification mirror stochastic strategies of cellular diversification. This will add a new economic analogy to the economic analogies of previous chapters.        

\section{Strategies to cope with uncertainty: a financial analogy}

\begin{figure}[t!]
  \begin{center} 
   \includegraphics[width=1.0\textwidth]{\chapterroot 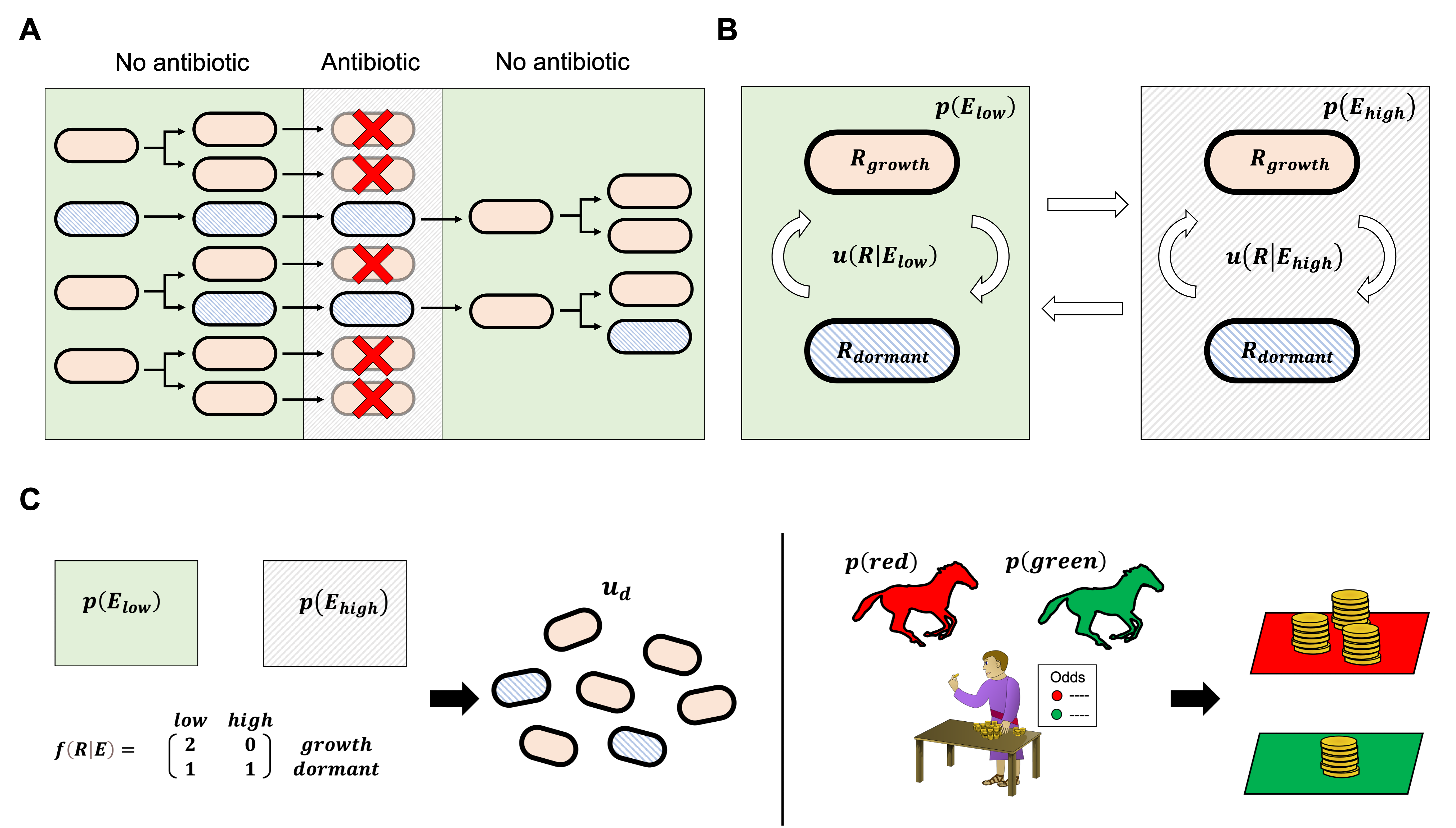}
   \caption{Bacterial persistence as an example of a cellular strategy to cope with uncertainty in environmental conditions. {\bf A)} Cells in a genetically identical population can display one of two distinct phenotypes that are associated with growth (pink cells) or dormancy (blue cells) in the absence of antibiotics. Only the dormant cells survive (persist) when exposed to antibiotics, and can transition back to the growth phenotype so that the population as a whole resumes growth in the absence of antibiotic. {\bf B)} In this simplified model of bacterial persistence, the strategy $u$ over two responses (phenotypes) $R_{\rm growth}, R_{\rm dormant}$ depends on environmental states $E_{\rm low}$ and $E_{\rm high}$, corresponding to low and high levels of the antibiotic, respectively. The occurrence of the states $E_{\rm low}$ and $E_{\rm high}$ is governed by probabilities $p(E_{\rm low})$ and $p(E_{\rm high})$, respectively. {\bf C)} The multiplicative rates $f(R|E)$ associated with phenotypes $R_{\rm growth}, R_{\rm dormant}$ depend on environmental conditions, so that $f(R|E)$ can be represented in matrix form. The resulting optimal strategy $u_d$ corresponding to the fraction of dormant cells in the population in turn depends on the probabilities of the environmental state $E$. An analogy with Kelly betting is illustrated on the right-hand side, where the probabilities of a horse winning a race, the odds provided by a bookmaker and the optimal betting strategy are identified with $p(E)$, $f(R|E)$ and $u(R|E)$, respectively, as displayed in Table~\ref{tab}.}
   \label{fig:persist}
  \end{center}
\end{figure}

We will use the topic of bacterial persistence as a recurring example throughout this chapter (Figure \ref{fig:persist}). When a clonal population of bacteria is exposed to an antibiotic, not all cells within the population are killed -- a small sub-population, although genetically identical to the rest, may nevertheless be in a distinct phenotypic state that is growth-dormant and resistant to treatment (Figure \ref{fig:persist}A). While the peers of this dormant sub-population previously grew well in the absence of antibiotic, upon exposure to treatment these growing cells are killed, and only the dormant cells (the persisters) remain alive. In turn, when the remaining persisters are transferred to an environment without antibiotic a large fraction is able to revert to the growing state, allowing the population as a whole to survive. Remarkably, in this subsequent phase of growth roughly the same small fraction of persisters is retained as before the treatment. Deterministic models based on short-term optimal growth cannot explain how part of a population adopts a slow-growing state: they would predict that each cell should adopt the growing phenotype in absence of antibiotics. Cells could have a mechanism to detect the presence of unfavorable environmental conditions and adopt the persister phenotype as a response, but there are several experimental observations not explained by such a mechanism~\cite{BalabanLeibler2004}: (1) a fraction of persisters exists prior to antibiotic treatment; and (2) not all cells, although genetically identical, adopt the persister phenotype. We will see that a more parsimonious description of persistence involves an optimization of long-term rather than short-term growth, which differs when environmental conditions fluctuate. 

Bacterial persistence is an example of {\em bet-hedging},
which more generally refers to the benefit of spreading resources across multiple behavioral phenotypes to reduce the risk associated with investing all resources into any single phenotype.
Returning to the example of bacterial persistence, a natural question one may ask is: what determines the precise fraction of persister cells (risk-avoiding, potentially low-reward phenotype) compared to growing cells (risky, potentially high-reward phenotype) within a given population? 
This question echoes a central question in financial investment: how should investors diversify their portfolio to maximize their capital in the context of uncertain returns? We will see that some of the same mathematical arguments of optimality under uncertainty can be used to analyze these two problems, showing how the optimal fraction of persisters is expected to depend critically on the probability to experience different environmental states. The terms of the analogy are presented in Table~\ref{tab}.

\begin{table}[t]
\begin{center}
\begin{tabular}{|l|l|l|}
\hline\label{tab}
{\bf Biology} & {\bf Gambling } & {\bf Finance} \\ 
\hline
Individual & Capital unit & Currency unit \\ 
Environment $p(E)$ & Race results $\p(x)$ & Market state \\
-- & Gambler & Investor \\
Phenotype decisions $u(R)$ & Bets $\B(x)$ & Investment strategy \\
Multiplicative rate $f(R,E)$ & Odds $\odds(x)$ & Immediate return \\
Environmental cue $P(S|E)$ & Side information $P(y|x)$ & Side information \\
Population growth rate $\Lambda$ & Long-term return $W$ & Long-term return \\
Extinction probability & Probability of bankruptcy & Probability of bankruptcy \\
Growth rate variance $\sigma^2$ & Growth rate variance $\sigma_W^2$ & Volatility \\
Population size $N_t$ & Capital $C_t$ & Capital \\ 
\hline
\end{tabular}
\end{center}
\caption{Analogy between bet-hedging in biological populations and diversification strategies in Kelly's gambling and finance. The common problem in each case is an uncertain environment that makes it impossible to anticipate which phenotype or investment is optimal for future growth. In finance, the ``population'' is constituted by the capital which is distributed across different options (different horses of a race or different stocks of a stock market). The main limitation of the analogy is that information is not processed centrally in biological populations but at the level of each individual, with therefore no equivalent to a gambler or investor. The notations are introduced in the main text for the biological problem and in Box 2 for the gambling problem.}
\end{table}

A pure bet-hedging strategy assumes the absence of any direct information on the current environmental state. Biologically, cells may sense signals or cues that encode varying degrees of information on their current environment.
For instance, in some populations, a larger proportion of persisters is found in nutrient-poor environments compared to nutrient-rich, implying a direct relationship between shifts in environment and switches between phenotypes. These sensing or signaling mechanisms can come with associated costs however, imparted by the investment of cellular resources in, for example, the gene expression machinery. Thus, optimal cellular behavior in the face of uncertainty may be expected to involve a trade-off between stochastic (e.g., bet-hedging) and deterministic (e.g., signaling) mechanisms that balance benefit to cost in a manner that depends on evolutionary context. 
Other trade-offs may also exist regarding reward versus risk associated with a particular cellular decision making strategy. Analogously, financial investors face trade-offs when using incomplete information on the current state of the market and developing an investment strategy based on the level of risk they are willing to incur. 

\section{Modeling cells growing in uncertain environments}

We begin with a simple model of persistence before introducing a more general framework. This simple model assumes that bacterial cells experience an alternation of low and high antibiotics environments and can adopt two physiological states, growing or dormant (Fig.~\ref{fig:persist}). The dormant cells are unable to replicate but persist in either high- or low-antibiotics environments while growing cells always divide when antibiotics are low in concentration but die when they are high. Mathematically, this is described by $f(R,E)$, the number of descendants of a cell with phenotype $R$ in environment $E$: $f(R={\rm dormant},E={\rm low}) = f(R={\rm dormant},E={\rm high}) = 1$, while  $f(R={\rm growing},E={\rm low}) = 2$ and $f(R={\rm growth},E={\rm high}) = 0$. In absence of sensing mechanism, we consider that the fraction of dormant cells, $u_d \equiv u(R={\rm dormant})$, is a fixed quantity that only possibly evolves on very long time scales. The population thus grows by a global factor $A_{\rm high}=f(R={\rm dormant},E={\rm low})u_d$ if the environment is high antibiotics and by a factor $A_{\rm low}=f(R={\rm dormant},E={\rm low})u_d +2f(R={\rm growing},E={\rm low})(1-u_d)$ if it is low antibiotics. Finally, the environment fluctuates randomly, with a probability $p_a$ to have high antibiotics and a probability $1-p_a$ to have low antibiotics. Over a large number $T$ of generations, a population therefore experiences in average $p_aT$ periods of high antibiotics and $(1-p_a)T$ periods of low antibiotics. As further explained below, the population size $N_T$ after $T$ generation is hence expected to globally grow as $N_t=(A_{\rm high})^{p_aT} (A_{\rm low})^{(1-p_a)T}N_0$. This corresponds to an exponential growth (or decay) of the form $N_T=e^{\Lambda T}N_0$ with a long-term growth rate $\L$ given by $\Lambda=p_a\ln u_d+(1-p_a)\ln(u_d+2(1-u_d))$.

Two bacterial populations which have different ``strategies'' $u_d$ will then have different growth rates $\L(u_d)$. The optimal strategy which maximizes $\L(u_d)$ is therefore when the probability $u_d$ to adopt the dormant state is 
\begin{equation*}
  u_d =\begin{cases}
    2p_a, & \text{if $0<p_a \leq 1/2$}.\\
    1, & \text{if $ 1/2 < p_a \leq 1$}.
  \end{cases}
\end{equation*}
The interesting case is when $p_a<1/2$, otherwise antibiotics is so often high that the population cannot grow. In this case, we find that a limited fraction of the population should be in the dormant state and that this optimal fraction depends on the frequency $p_a$ at which high antibiotics occurs. 

This example can be extended to an arbitrary number of environmental states $E$ and phenotypic (response) states $R$ and to the presence of cues collected from the environment. In general, the states and cues may take discrete (as in the above example) or continuous values. 
The ``strategy'' of a cell may then be described by its probability $u(R)$ to adopt a particular phenotype $R$. This strategy depends on the environment if some signal $S$ is perceived, in which case the strategy takes the form of a conditional probability $u(R|S)$ satisfying
\begin{equation*}
    \sum_R u(R|S) = 1, \quad {\rm with} \quad u(R|S) \geq 0 
\end{equation*}
for each possible signal $S$. For the example of bacterial persistence, $u(R={\rm dormant}|S)$ may be the fraction of cells adopting a dormant phenotype within the population of cells with intracellular antibiotics concentration $S$. The fraction of growing cells would then be given by $u(R={\rm growing}|S) = 1 - u(R={\rm dormant}|S)$. By comparison, Figure \ref{fig:persist}B illustrates a model where $u(R|E)$ depends directly on the environmental state $E$. In finance, $u(R|S)$ would correspond to the fraction of the capital that an investor allocates to asset $R$ when receiving incomplete information $S$ on the current market state $E$. More generally, we may also consider that the probability to adopt a phenotype $R_t$ at time $t$ depends on the phenotype $R_{t-1}$ adopted at time $t-1$ by the cell or its parent, which would be described by $u(R_t|S,R_{t-1})$ or $u(R_t|S_t,R_{t-1})$ to indicate that the signal $S_t$ is obtained at time $t$.

The model also needs to specify the temporal dynamics of the environment and the relation between $S$ and $E$. The simplest assumption is that successive environmental states are uncorrelated, and occur with probability $p(E)$ and that signals are derived from a conditional probability $p(S|E)$, as illustrated in Figure \ref{fig:persist}B where $p(S|E) = \delta(S|E)$ is equivalent to $S\equiv E$. This is sufficient to demonstrate bet-hedging or discuss the value of signaling and in the examples below we therefore make this simplifying assumption by default. More generally, to address issues of inheritance where $R_t$ depend on $R_{t-1}$, we may assume a discrete-time Markov process where the state of the next environment depends only of the previous one, with transition probabilities $p(E_t|E_{t-1})$ where $E_t$ denotes the state of the environment at time $t=1,2\dots$. 
Even more generally, we may also want to account for the feedback that the population exerts onto its environment and consider that $E_t$ depends on the size and composition of the population.

Finally, we need to specify the dynamics of the population itself. Between time points $t$ and $t+1$, a cell adopting phenotype $R$ in the context of environment $E_t$ either dies or survives and may additionally produce offsprings. This is summarized by a quantity $f(R,E_t)\geq 0$ that indicates the mean number of descendants at time $t+1$ of an individual with phenotype $R$ in environment $E_t$ (possibly including the individual itself). Given that $u(R|S_t)$ denotes the fraction of cells or probability of the organism adopting phenotype $R$ based on sensed state $S_t$, a population is therefore expected to globally increase (or decrease) in size by a factor 
\begin{equation}
A_t=\sum_R f(R,E_t)u(R | S_t)
\end{equation}
that depends both on the strategy $u$ and the current environmental state $E_t$. This factor $A_t$ is a stochastic variable as it depends on the stochastic variables $E_t$ and $S_t$. More explicitly, if $N_t$ denotes the size of the population at time $t$, this size will increase or decrease to $N_{t+1} = A_tN_t$ at time $t+1$ (in average). We can in this way account for the dynamics of population growth and then ask what is an ``optimal'' strategy $u(R|S)$ that leads to, for example, the largest population size over a given time interval. Compared to the deterministic setting, however, this is not yet a well-formulated problem as the population size varies with time and therefore generally depends on the particular sequence of environments $E_0,\dots,E_t$, which is in turn stochastic. Thus, we need to extend the concept of optimality to the stochastic regime. We examine this question in the next section.

\section{Optimization in uncertain environments}


In the previous section, we used notation $A_t$ to denote the fractional increase or decrease in population size given that strategy $u(R|S_t)$ is adopted in environment $E_t$. An alternate name for this quantity is the instantaneous growth rate. It follows from recursion that, given an initial population size of $N_0$ at time $t=0$, the population size at time $t$ is given by
\begin{equation}\label{eq:mf}
N_t = A_t A_{t-1} \cdots A_1 N_0
\end{equation}
where $A_t$ depends on the environmental state $E_t$ and is therefore a stochastic variable when the environment varies stochastically.
Here the choice of an objective criterion is fundamentally linked to the time scale at which growth is considered.

\subsection{Long-term versus short term optimization}

At the shortest time scale, maximization of population growth over a single time step corresponds to adopting the distribution $u(R|S_t)$ that maximizes the arithmetic mean $\E[A]$, where $A$ denotes the random variable whose realization at time $t$ is $A_t$ (Box~1). This maximum is typically achieved by a population where all individuals adopt the same optimal phenotype -- the phenotype $R$ maximizing $\E[f(R,E_t)u(R | S_t)]=\sum_{E,S}P(S|E)P(E)f(R,E)u(R | S)$. In the example of persistent cells, this strategy would correspond to having all cells in a growing state if the most likely environment is an absence of antibiotics. This strategy is extremely risky if these growing cells cannot survive an episode of antibiotics, which would therefore lead to extinction of the population. Taking into account the rare but important events of high antibiotics concentration requires taking a long-term perspective. Remarkably, in the long-term the problem becomes effectively deterministic due to the law of large numbers. The best known example of a law of large number applies to the sum $A_1+\dots+A_t$ of $t$ random variables $A_i$, which almost certainly behaves as $t\E[A]$ as $t\to\infty$. 
Here, the problem involves a {\it product} of random variables and a similar but different law of large number applies: the product $A_1\times\dots\times A_t$ does not typically behave as $(\E[A])^t$ but instead as $\exp (t\E[\ln A])$ where $\E[\ln A]$ is known as the geometric mean (Box~1).
This corresponds to the intuition that population size typically grows exponentially in the long run, $N_t\sim e^{\L t}N_0$, with a well-defined long-term growth rate 
\begin{equation}
\label{Lambda}
\L=\E[\ln A]= \sum_{E,S} p(S | E) p(E) \ln \left(\sum_R f(R,E)u(R | S)\right),
\end{equation}
that is predictable despite the stochasticity of the environment. 

Biologically, therefore, maximizing the geometric mean is equivalent to maximizing the long-term growth rate of the population. This is the relevant measure of fitness in the long-term from an evolutionary point of view, because of two populations with growth rates $\L_1$ and $\L_2$, the one with $\L_1>\L_2$ will almost certainly exponentially outnumber the other. 

\begin{MathDetailblock}[title=Arithmetic versus geometric mean and logarithmic utility functions,float=h]
\label{box:geometric}
Additive random processes are governed by the law of large numbers: the sum of many random variables scales with their arithmetic mean. In finance and biology, returns are compounded and growth is a multiplicative process. This is fundamentally different: the typical outcome is no longer described by the arithmetic mean but by the geometric mean~\cite{Redner1990}. A simple example illustrates this difference. Imagine a succession of environments in which the population either doubles or is reduced by $2/3$, with same probability. This corresponds formally to a population size increasing as $N_t=A_t\dots A_1 N_0$ where $A_t=2$ (doubling) with probability $1/2$ and $A_t=1/3$ ($2/3$ dying rate) with probability $1/2$. The arithmetic mean is $7/6$ which is $>1$ and suggests that the population will grow. But as each outcome has the same probability, the typical growth over $t$ generation is actually given by $2^{t/2}(1/3)^{t/2}=e^{t\L}$ with $\L=(1/2)\ln(2/3)$ which is $<1$: the population will in fact most likely go extinct. Mathematically, taking the log turns the product into a sum to which the central limit theorem applies. More intuitively, the arithmetic mean is dominated by very rare events. Historically, the importance of the geometric mean for estimating risk was first understood by Daniel Bernoulli in the context of games~\cite{Bernoulli1738,Stearns2000}. Later, it has been the subject of many debates in finance~\cite{Bernoulli1738}, reflecting the fact that alternative {\it utility functions} over which to optimize may be more appropriate when considering a short temporal horizon or when accounting for different degrees of risks.
\end{MathDetailblock}

The simple example of persistence that we introduced previously illustrates well how maximizing the long-term growth rate is different from optimizing the instantaneous growth rate. The arithmetic mean $\E[A]$ is indeed maximized by $u_d=0$ when $p_a<1/2$, which leads to certain extinction unless $p_a=0$. 
This remains true for general models including multiple environmental states and sensing that conveys information about the environment through conditional probability $p(S|E)$. Using the long-term growth rate $\Lambda$ as a measure of fitness, it is then possible to quantify the value of information $S$ by comparing the optimal growth rate that can be achieved in presence of $S$ to that in its absence. Remarkably, for special limits of the model, corresponding to Kelly's horse-race model (Box~2), this value is given by some of the same quantities that appear in Shannon's theory of communication (Box~3). 

\begin{figure}[t!]
  \begin{center} 
   \includegraphics[width=0.5\textwidth]{\chapterroot 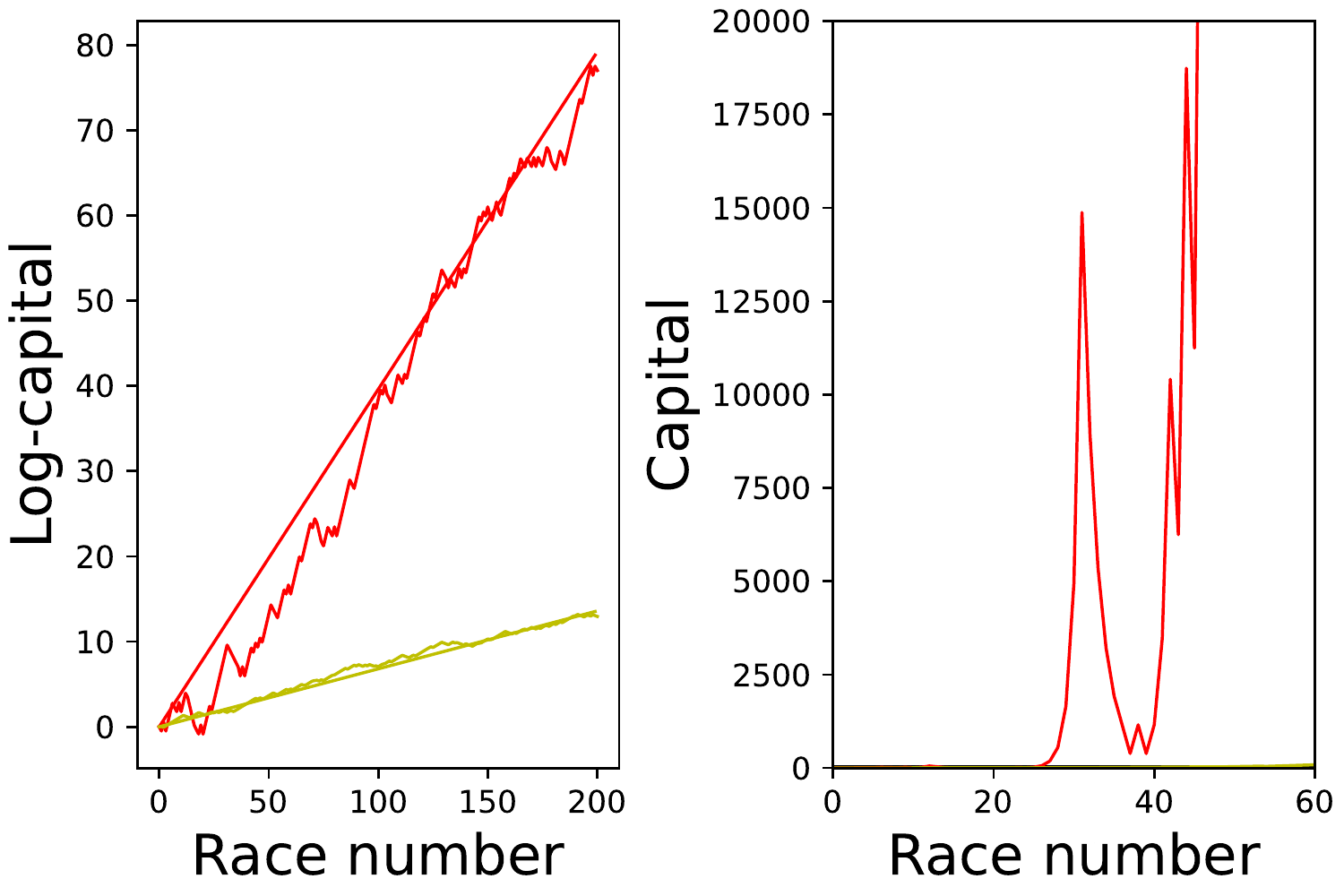}
   \caption{Evolution of the log-capital (left plot) or of the capital itself (right plot) as a function of the number of races for Kelly's optimal strategies (red curve) and for a non-optimal strategy (yellow curve). On left plot, the straight lines have the slope of the corresponding growth rate for each strategy. Note that the fluctuations in Kelly's strategy can in fact be quite large, when plotted in normal scale instead of log-scale.}
   \label{fig:Kelly}
  \end{center}
\end{figure}

\begin{MathDetailblock}[title=Kelly's model]\label{box:kelly}
In 1956, Kelly \cite{Kelly1956}  extended the work of Shannon on communication to the field of gambling. This classic model has important implications for investment strategies in finance and beyond. In the context of biology, Kelly's paper led to a clarification of the notion of value of information which is described in Box 3. 	

Let us recall the basic elements of Kelly's horse race. 
The odds paid by the bookmaker when the horse $x$ wins is $\odds(x)$, and
the probability for this to happen is $\p(x)$. A gambler can
distribute his/her bets on the different horses, and $\B(x)$ is the
fraction of the bet set on horse $x$. 
Thus, a \emph{strategy} of the gambler is defined by 
a vector of bets $\mathbf{b}$ of $M$ components $\B(x)$. At every race, the gambler invests his/her entire capital on all horses, so that 
$\sum_{x=1}^{M} \B(x) = 1$, always betting a non-zero amount on all horses.
Since no bet is zero, there is a well-defined vector of the inverse of the odds paid by the bookmaker denoted $\mathbf{r}$.
When the odds are \emph{fair}, the bookmaker does not keep any of the invested capital and as a result $\sum_{x=1}^M \R(x) = 1$. 

At each time $t$, one horse, which we call $x$, wins with probability $\p(x)$. As a result, the capital at time $t+1$ is updated according
$C_{t+1} = \dfrac{\B_x}{\R_x} C_t$. 
As explained previously, this multiplicative process is best studied by considering instead the log of the capital, $\lp (t) \equiv \log C _t$, which satisfies the assumptions of the law of large numbers when races are independent. In these conditions, $\lp (t) \equiv \log C _t$ converges on long times towards the growth rate $W(\mathbf{b}, \mathbf{p})$ where 
\begin{equation}
W (\mathbf{b}, \mathbf{p})=  \sum_x \p(x) \log \odds(x) \B(x).  
\end{equation}

This growth rate can be rewritten using an information theoretic measure between two probability distributions, $\mathbf{p}$ and $\mathbf{q}$, called the Kullback-Leibler divergence and defined by 
\begin{equation}
\dkl( \mathbf{p},\mathbf{q} ) = \sum_x \p(x) \log \frac{\p(x)}{\q(x)}.
\end{equation}
One can show that this quantity is a non-negative measure between the two probability distributions.
With this notation, the growth rate can be rewritten as 
\begin{equation}
W (\mathbf{b}, \mathbf{p})=  \dkl \left( \mathbf{p} \Vert \mathbf{r} \right) - \dkl \left( \mathbf{p} \Vert \mathbf{b} \right),  
\end{equation}
It follows from this equation that the strategy $\mathbf{b}^* = \mathbf{p}$ is optimal. This strategy, known as Kelly's strategy or proportional betting, overtakes any other strategy in the long-term as illustrated in Fig.~\ref{fig:Kelly}. 

This formulation shows that the growth rate is the difference between the distance of the bookie’s estimate from the true distribution and the distance of the gambler’s estimate from the true distribution. Hence, the gambler makes money if they have a better knowledge of the winning probabilities than the bookie. The optimal long term growth rate is the positive quantity : 
\begin{equation}\label{eq:W}
W^* (\mathbf{b}, \mathbf{p})=  \dkl \left( \mathbf{b} \Vert \mathbf{r} \right).  
\end{equation}



Kelly's horse race model is formally a particular case of the model introduced in the main text when considering that one, and only one phenotype $R=R(E)$ can grow in any given environment $E$, such that $f(R,E)=f(E)$ if $R=R(E)$ and $0$ otherwise. Horses $x$ may then be interpreted as both the environments $E$ and their associated phenotypes $R(E)$ so that $u(R)=\B(x)$ and $f(E)=\odds(x)$. 
In biology, but also in finance where $R$ is interpreted as an asset, there is generally no one-to-one correspondence between environments $E$ and phenotypes $R$ and multiple phenotypes (assets) may grow (have non-zero return) in any given environment. 
The optimal strategy is then no longer necessarily proportional betting as illustrated in the example of persistence presented in the main text and as also shown in Ref.~\cite{Pugatch-preprint}.

 
\end{MathDetailblock}

\begin{MathDetailblock}[title=The value and cost of information for growing populations,float=h]\label{box:info}

To see how uncertainty may be quantified by Shannon entropy, first consider a model where $f(R,E)$ is non-zero only for one phenotype $R$ best adapted to each particular environment $E(R)$. As seen in Box~2, the optimal strategy in the long-term is proportional betting, $u(R(E))=p(E)$. To quantify the cost of uncertainty, it must be compared to a situation where full information is available, in which case all the cells can systematically adopt the optimal phenotype, leading to an ideal growth rate $\L^{**}=\sum_E p(E)\ln f(R(E))$. The cost of uncertainty is $\L^{**}-\L^*=-\sum p(E)\ln p(E)$, which is nothing but the Shannon entropy of the environment $H(E)$. This has a simple interpretation: the more unpredictable the environment, the larger its entropy and the lower the maximal growth rate of the population.

The reasoning can be extended to the presence of partial information, modeled by $p(S|E)$. The case of perfect information is indeed the limit case where $S=E$. The optimal strategy with partial information is a generalization of proportional betting that takes into account $S$ and the difference of growth rate is now given by the mutual information $I(S,E)$ (Exercise~\ref{ex:partialinfo1}). The mutual information is minimal when the signal $S$ is uncorrelated to $E$, in which case $I(S,E)=0$, and maximal in presence of perfect information, in which case $I(S,E)=H(E)$~\cite{CoverThomas2006}.

These results were first derived by Kelly~\cite{Kelly1956}. They have been generalized to more general forms of $f(R,E)$ as well as to more general environmental processes in the context of financial investment in which case the cost of uncertainty and value of information are no longer equal but bounded by information theoretic quantities~\cite{CoverThomas2006,RivoireLeibler2011}. This is illustrated in Exercise~\ref{ex:partialinfo2} with an extension of the model of persistence presented in the main text.

Information is generally costly as it implies producing and operating an accurate sensor, which may come at the expense of growth rate. Taking into account this cost introduces a trade-off between the cost and value of information that may justify an imperfect sensor, or even explain an absence of sensor (Exercise~\ref{ex:partialinfo1}). This trade-off has for instance been invoked to explains that bacteria subject to infrequent periods of antibiotics evolved to stochastically switch their phenotype rather than to sense the presence of antibiotics~\cite{KussellLeibler2005}.

While the problems of information processing in biology and in finance share many analogies, it is also important to recognize an important difference: in biology, information processing is distributed at the level of each cell, which may perceive different signals, while in finance, information is processed by an investor who centralizes the information. The value of information is bounded by information theoretic quantities only in the second case, or more generally when the same common information is available to all the cells~\cite{RivoireLeibler2011}. If information processing is stochastic at the single cell level, the value of information is effectively higher (Exercise~\ref{ex:partialinfo3}).

\end{MathDetailblock}

\subsection{Trade-offs at intermediate time scales}

So far we considered two extreme limits of immediate and infinite time scales under one important assumption: the population is always large enough to escape extinction. Eq.~\eqref{eq:mf} is indeed valid only for large $N_t$ and does not apply anymore when $N_t\sim 1$, in which case the population size is subject to stochastic effects, called demographic noise in population biology. In our analogy with finance, the eventuality of $N_t=0$ with no possible recovery corresponds to a risk of bankruptcy. 

When considering long time scales, a population with $\L>0$ will either become extinct or grow exponentially. In this later case, demographic noise is eventually negligible and our approach valid. At intermediate time scales, however, population sizes $N_t$ may deviate substantially from $N_0e^{\L t}$ predicted by exponential growth, and may become extinct ($N_t=0$) as a result. To quantify these deviations, note that for
the model defined in the main text where there are no correlations of the instantaneous growth rate $A_t$, the central limit theorem imposes that the quantity
\begin{equation}
\label{Delta_t}
\Delta_t =\frac{1}{\sigma \sqrt{t}} \left( \log \frac{N_t}{N_0} - t \Lambda \right),
\end{equation}
converges on long times towards a Gaussian distribution of unit variance, where $\sigma$ is the standard deviation of the instantaneous growth rate. It follows from this property that  
\begin{equation}
\label{risk}
\sigma^2 = \frac{1}{t}  {\rm Var} \left( \log \frac{ N_t}{N_0}\right),
\end{equation}
measures the deviation from exponential growth. This quantity is therefore a natural measure of risk, known in finance under the name of volatility. 
To understand at which time scale this risk is important, we consider Eq. \ref{Delta_t}, assuming $\Delta_t$ is of the order one. Risk will be important, when the term associated with fluctuations, which is of the order of $\sigma \sqrt{t}$ will be larger than the term associated with exponential growth, which is $t\Lambda$. This will happen when $t \ll (\sigma/\Lambda)^2$: the risk is relevant at intermediate time scales, long-enough for the central limit theorem to apply but not too long for deviations from exponential growth to become negligible. 

This measure of risk has well known drawbacks in finance : it is symmetrical with respect to losses and gains, which does not conform to the intuitive notion of risk, and furthermore typical fluctuations are often non-Gaussian. Nevertheless, the volatility is still an important notion in the study of optimization of portfolios \cite{Markowitz1952}. In this context, Markowitz introduced plots of the volatility $\sigma$ as a function of the mean growth rate, which define the so-called ``efficient frontier''. This representation illustrates graphically a fundamental trade-off that exists between the maximization of the mean return and the minimization of the variance (or risk). The point of zero volatility is a risk-free strategy, which corresponds to dormant states in biology. 

\begin{figure}[t!]
  \begin{center} 
   \includegraphics[width=0.5\textwidth]{\chapterroot 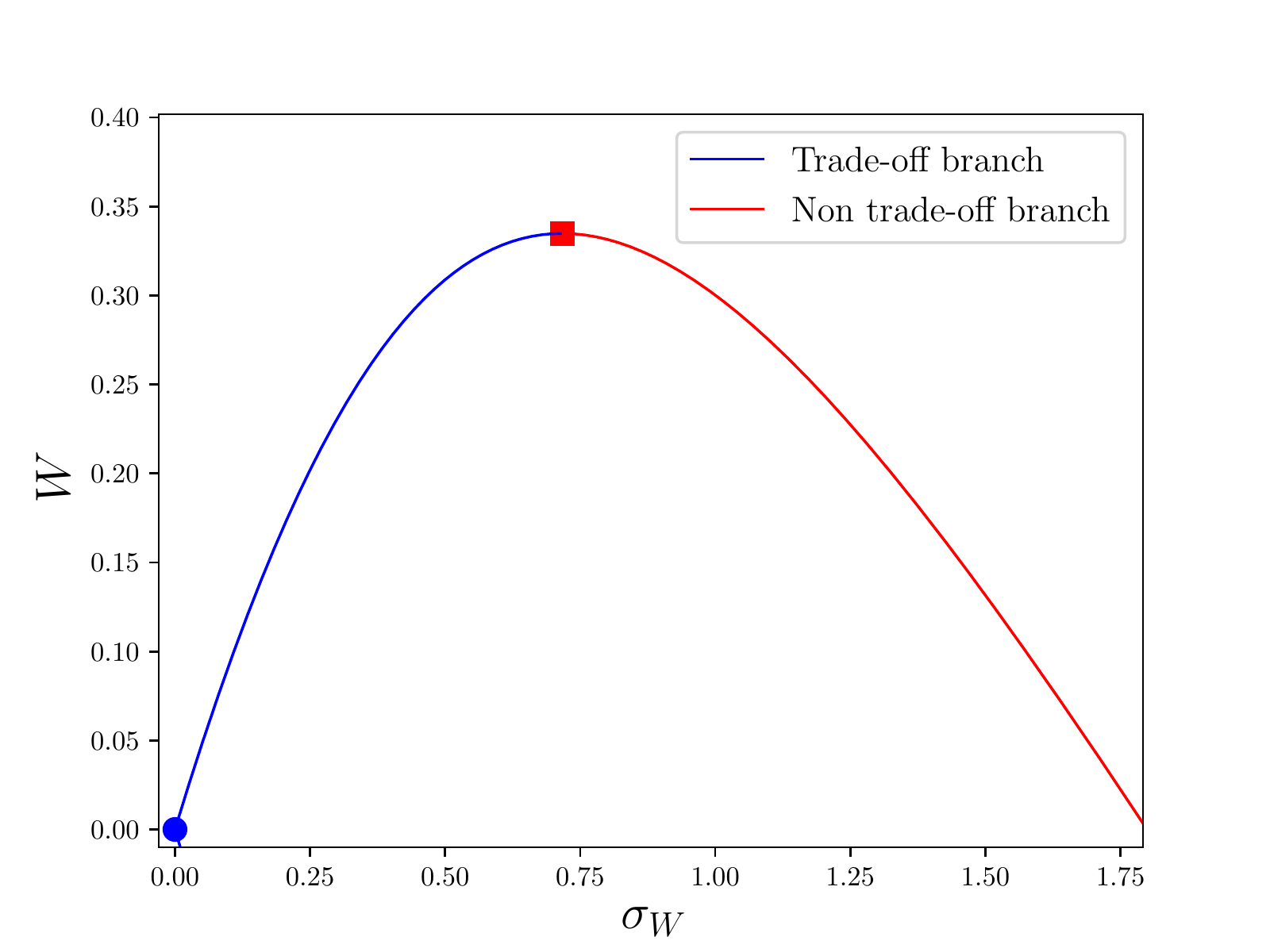}
   \caption{Pareto diagram showing the growth rate $W$ versus the fluctuations measured by the standard deviation of the growth rate $\sigma_W$ (which is the volatility for this model) in the simple case that only two horses are present. The curve can be obtained by varying a risk aversion parameter $\alpha$, which enters in the definition of an objective function (see Exercise~\ref{ex:pareto} for details). The point of maximum growth rate (red square) corresponds to Kelly's strategy and divides a trade-off branch (blue solid line) from a non-trade-off branch (red solid line) (adapted from \cite{DinisLacoste2020}).}
   \label{fig:2horses}
  \end{center}
\end{figure}

This trade-off is naturally present in Kelly's model introduced in Box~2. Indeed, Kelly's strategy is based on the maximization of the long-term growth rate, but at intermediate times the capital can deviate significantly from the expected exponential growth
as shown in figure \ref{fig:Kelly}. Prominent economists, such as Samuelson, strongly opposed the use of Kelly's criterion in finance precisely for that reason \cite{Samuelson1971}. 
In practice, however, investors can mitigate this risk by using Kelly's criterion for only a fraction of the bets \cite{MacLeanZiemba2011}. The resulting strategy has reduced fluctuations, and at the same time, a reduced growth rate. Another consequence of the trade-off is that the risk near the optimal strategy (Kelly's strategy) can be reduced significantly provided one is ready to sacrifice a small amount of growth rate, an important lesson for gamblers and investors. 
In order to build systematically improved gambling strategies with a reasonable amount of risk in Kelly's model, one can introduce an objective function that is a linear combination of the growth rate with the volatility of Kelly' model, $\sigma_W$, weighted by a risk aversion parameter $\alpha$ \cite{DinisLacoste2020}. The method is illustrated in Exercise~\ref{ex:pareto} for the two-horse version of Kelly's model. By optimizing this objective function, one builds the Pareto diagram shown in Fig.~\ref{fig:2horses} when varying the parameter $\alpha$.

A general inequality characterizes this trade-off mathematically for an arbitrary number of horses. 
For Kelly's gambling model with fair odds
defined in the box \ref{box:kelly}, it has the form 
\begin{equation}
\sigma_W \ge  \frac{W}{\sigma_q},
\end{equation}
where $\sigma_W$ is the volatility of Kelly's model, $W$ the average growth rate (the equivalent of $\Lambda$) and $\sigma_q$ is the standard deviation of a distribution, $\q(x)$ defined by $\q(x)=\R(x)/\p(x)$. This distribution compares the probability of races outcomes described by $\p(x)$ with the risk-free strategy described by $\B(x)=\R(x)$, for which $\sigma_W=W=0$ \cite{DinisLacoste2020}. Recently, a similar bound has been derived for other well-known financial models such as the Black-Scholes and the Heston models \cite{ZiyinMasahito2022}.

Let us now illustrate the implications of this trade-off for a biological population using a simple bet-hedging model with only two phenotypes. Individuals in the population can switch from phenotype $A$ to phenotype $B$ with a transition probability $\pi_1$, and with probability $\pi_2$ from $B$ to $A$, assuming no sensing. The population grows in an environment that fluctuates between two values 1 and 2.
We denote the population vector, which describes the number of individuals 
in each phenotype  at a given time $t$ by 
${\bf N}(t)=(N_A(t),N_B(t))^T$, where $T$ denotes the transpose. 
The subpopulation of individuals with phenotype $A$ grows when placed in the environment $i$ with the growth rate $k_{Ai}$, while the other subpopulation with phenotype $B$ grows with rate $k_{Bi}$. The population is assumed to be large, there is no population noise, the dynamics of the system is deterministic in each separate environment. 
The population dynamics of the model can be described by the vector equation :
\be
\frac{d}{dt} {\bf N}(t)= M_{S(t)} {\bf N}(t),
\label{eq_ev}
\ee
with  matrices
\be
M_{S_1} = 
\begin{pmatrix}
k_{A1}-\pi_1 & \pi_2 \\
\pi_1 & k_{B1} -\pi_2 
\end{pmatrix} 
\, {\rm and} \, \, \,
M_{S_2} = 
\begin{pmatrix}
-\pi_1 + k_{A2} & \pi_2 \\
\pi_1 & k_{B2}-\pi_2 
\end{pmatrix}. 
\ee

The finite time averaged population growth rate is defined as
\be
\label{Lambda_t}
\Lambda_t=\frac{1}{t} \ln \frac{N(t)}{N(0)},
\ee
in terms of the total population 
$N(t)=N_A(t)+N_B(t)$,
and the long term population growth rate is 
\be
\Lambda= \lim_{t \to \infty} \Lambda_t.
\ee

\begin{figure}[t!]
  \begin{center} 
   \includegraphics[width=0.5\textwidth]{\chapterroot 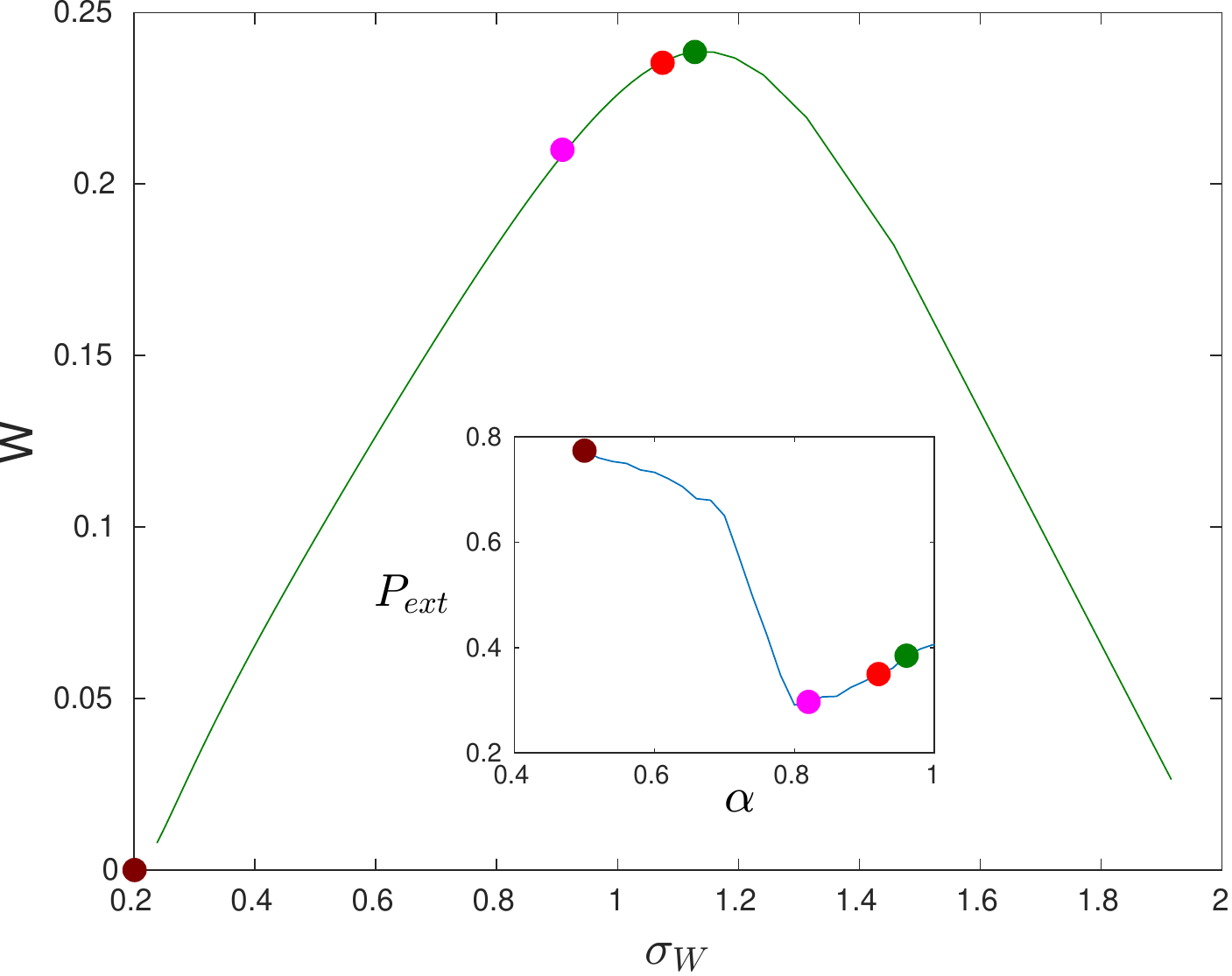}
   \caption{Pareto diagram showing the population growth rate versus the fluctuations of that growth rate in a simple model of a biological population evolving in a stochastic  environment with no sensing according to Eq. \ref{eq_ev} \cite{DinisLacoste2022}. In this figure, the time scale of environment fluctuations is comparable to that of phenotypic fluctuations. The inset shows the probability $P_{ext}$ that the population goes below a certain extinction level versus the risk aversion parameter $\alpha$ which measures the distance along the Pareto plot. Colored bullets represent different points on the Pareto front (adapted from \cite{DinisLacoste2022}). }
   \label{fig:Pareto-bio}
  \end{center}
\end{figure}

This optimal long term growth rate $\Lambda$ can be obtained analytically in this model~\cite{HuftonMcKane2016}, but approximations are needed  
to evaluate the fluctuations of the growth rate, which is the equivalent of the volatility  $\sigma^2$ of Eq.~\ref{risk} \cite{DinisLacoste2022}. 
One can then study the trade-off that exists between the average growth of the population (either measured instantaneously or over a long time) and the fluctuations of the growth rate, using the same Pareto plot used for Kelly's model in figure \ref{fig:2horses}.
This ``efficient frontier'' is shown in Fig.~\ref{fig:Pareto-bio}, and as in the case of Kelly's model, in the region of fast growth rate, it is advantageous for a population to trade growth for less risky fluctuations. In this model, $\sigma^2$ correlates with the probability that the population $N(t)$ goes below a certain threshold, where the population is considered as extinct. The probability of extinction is not monotonic along the Pareto front, which explains why in the region of low growth rate, it is more advantageous to prioritize instead the increase the growth rate to avoid extinction. 

In the context of ecology, besides the probability of extinction, a quantity of interest is the chance for a population to grow from rarity in the presence of other species. In agreement with the above trade-off, it was found that this chance can not be predicted only from the mean growth rate, and that the mean growth rate and its variance should be both used for such a prediction \cite{Pande2022}.
In summary, the similarity of the Pareto plots (called efficient frontier in finance) obtained in Kelly's model and in models of biological populations in fluctuating environments \cite{DinisLacoste2022}, and evidences from various works in ecology, suggest that the trade-off discussed here is broadly applicable in various fields ranging from biology and ecology to economics.

\section{Strategies in correlated environments}

So far we considered two time scales: the time scale at which phenotypic changes occur and at which instantaneous growth is defined ($t=1$ in our discrete-time model, which may be taken to correspond to one generation), and the longer time scale $t\sim (\s/\L)^2$ beyond which population growth is effectively exponential, with growth rate $\L$. We saw that the choice of an optimization criterion depends fundamentally on the time horizon relative to these time scales.

Additional time scales are relevant when environmental states are correlated in time, for instance through a Markov chain $P(E_t|E_{t-1})$. This is for instance the case if conditions of high nutrient or high stress extend over several generations. As a consequence, strategies $u(R_t|S_t,R_{t-1})$ that depend on past internal states $R_{t-1}$ in addition or instead of externally driven signals $S_t$ may become valuable, since the fact that phenotype $R_{t-1}$ survived in environment $E_{t-1}$ indirectly carries information on the current environment $E_t$. We may then recognize that $R_t$ plays two distinct roles: on one hand, it determines survival and growth via $f(R_t,E_t)$ and, on the other, it provides information to determine the next state $R_{t+1}$ via $u(R_{t+1}|S_{t+1},R_t)$. This corresponds to the fundamental distinction between phenotype and genotype in biology: the genotype $\g$ is what is transmitted from one generation to the next while the phenotype $\phi$ is what determines instantaneous growth. Formally, $R_t=(\phi_t,\g_t)$ with $f(R_t,E_t)=f(\phi_t,E_t)$ and $u(R_t|S_t,R_{t-1})=u(R_t|S_t,\g_{t-1})$, by definition of $\phi_t$ and $\g_t$. The ``central dogma'' of molecular biology states that information flows from the genotype to the phenotype but not reciprocally, which corresponds here to assuming that $u(\phi_t,\g_t|\g_{t-1})$ factorizes as $d(\phi_t|S_t,\gamma_{t-1})h(\g_t|\g_{t-1})$, where $d(\phi_t|S_t,\gamma_{t-1})$ may be interpreted as a developmental kernel and $h(\g_t|\g_{t-1})$ as an inheritance kernel, with no dependence on $S_t$ (no Lamarckism). The mathematical framework that we introduced can be used to study to which extent this particular decomposition is indeed a good ``strategy''~\cite{RivoireLeibler2014}. The answer generally depends on the nature and amplitude of the environmental fluctuations. 

\begin{figure}[t]
  \begin{center} 
   \includegraphics[width=0.9\textwidth]{\chapterroot 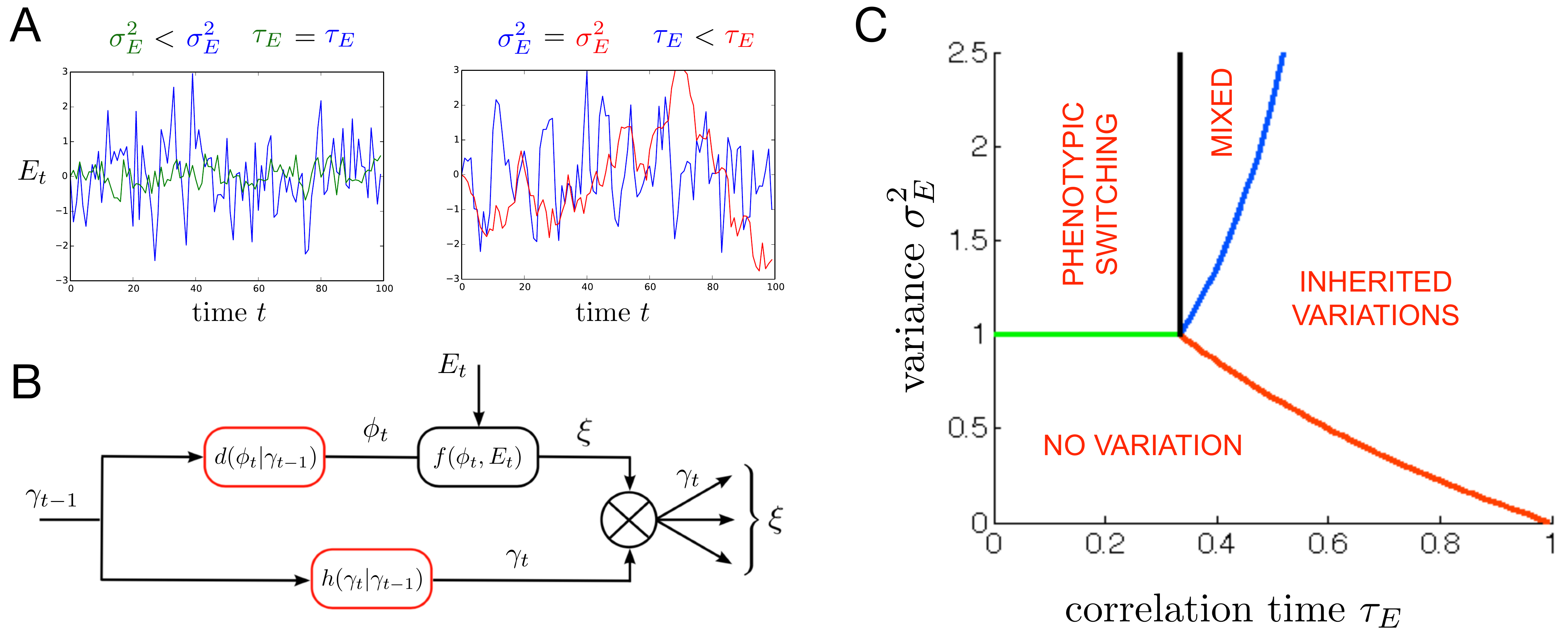}
    \end{center}
   \caption{Example of optimal strategies in correlated environments -- {\bf A.} We consider here a model where the environment $E_t$ is a continuous variable following a Gaussian process, $P(E_t|E_{t-1})=\exp(-(E_t-aE_{t-1})^2/(2\sigma_X^2))/(2\pi\sigma_X^2)^{1/2}$ with two parameters $a$ and $\sigma^2_X$ that control the overall amplitude of the fluctuations $\s_E^2=\s_X^2/(1-a^2)$ and their time scales $\tau_E=-1/\ln a$, as illustrated by the different time series. {\bf B.} An individual inherits a genotype $\gamma_{t-1}$ which determines its phenotype $\phi_t$ with probability $d(\phi_t|\gamma_{t-1})=\exp(-(\phi_t-\gamma_{t-1})^2/(2\sigma_D^2))/(2\pi\sigma_D^2)^{1/2}$ where $\sigma_D^2$ thus represents phenotypic noise. $\gamma_{t-1}$ also determines the genotype $\gamma_t$ of the progeny with probability $h(\gamma_t|\gamma_{t-1})=\exp(-(\gamma_t-\gamma_{t-1})^2/(2\sigma_M^2))/(2\pi\sigma_M^2)^{1/2}$ where $\sigma_M^2$ thus represents mutational noise. The number $\xi$ of offsprings is a random variable whose mean $f(\phi_t,E_t)=k\exp(-(\phi_t-E_t)^2/2)$ depends on the phenotype $\phi_t$ as well as the current state $E_t$ of the environment. A population of such individuals grows with a long-term growth rate $\L$ that can be computed analytically~\cite{RivoireLeibler2014}. {\bf C.} Values of $\sigma_D^2$ and $\sigma_M^2$ that optimize $\L$ define four phases as a function of the environmental parameters $\tau_E$ and $\sigma_E^2$. For nearly constant environments, the optimal strategy is to maintain constant phenotypes ($\sigma_D^2=0$) and genotypes ($\sigma_M^2=0$) (``no variation''). For strongly varying but poorly correlated environments, the optimal strategy is to introduce phenotypic variations ($\sigma_D^2>0$) but no genotypic mutations ($\sigma_M^2>0$) (``phenotypic switching''). For highly correlated environments, the optimal strategy is instead to introduce genotypic mutations ($\sigma_M^2>0$) while canalizing the phenopype ($\sigma_D^2=0$) (``inherited variations''). A phase also exists where both types of variations are beneficial (``mixed''). This model thus identifies environmental variations for which bet-hedging (phenotypic switching) is expected to evolve, namely variations of environmental of sufficient amplitude but with limited temporal correlations across generations.
   \label{fig:correnv}}
 
\end{figure}

Similarly, the model can be analyzed to understand the conditions under which it is advantageous to introduce phenotypic variations that are not transmitted -- as in bet-hedging -- versus genotypic variations that are transmitted -- as with genetic mutations. Stochasticity may indeed be introduced either in the mapping from $\g_{t-1}$ to $\phi_t$ or the mapping from $\g_{t-1}$ to $\g_t$, or in both of them -- a problem with no equivalent in finance. This is illustrated in  Fig.~\ref{fig:correnv} with a simple solvable model showing how the optimal strategy depends on the nature of the fluctuations of the environment. In particular, bet-hedging strategies where stochasticity is purely phenotypic are found to be optimal for environmental fluctuations of sufficient large amplitude but low temporal correlations from one generation to the next. 

Historically, the notions of genotype and phenotype were introduced much before the molecular mechanisms that underlie them were uncovered. In general, the genotype, defined as inherited information, should not be confused with the notion of genetic information: along with DNA, a range of epigenetic states, including metabolic states, are also transmitted from cell to cell which represent genotypic information. 
In other words, the physiological state of a cell, which we analyzed in most of this book from the standpoint of a phenotype determining current growth, may also represent valuable genotypic information for future generations.

\section{Perspectives}



We presented optimal strategies that biological populations may exploit for coping with uncertain environments and drew analogies with problems of gambling and financial investments. Optimality assumes a measure of performance which, however, is not readily defined when environments are changing stochastically. In particular, the time scale over which the problem is considered is critical. This difficulty has led to multiple debates over the concept of fitness in biology which partly mirror those over the concept of utility in economics.

While the analogy with finance is instructive, its limitations should also be kept in mind. Most importantly, the states that individuals of a biological population adopt are not centrally controlled by a gambler or an investor. This raises a question that is absent in finance but central in evolutionary biology: is a strategy that is optimal for the population but detrimental to some of its members -- as for instance the persister cells that ``sacrifice'' their current growth for the sake of future growth -- evolutionary stable? A strategy that is optimal for a population may indeed never be achieved through evolution as natural selection at the individual level may favor non-cooperating individuals -- an issue known as a ``conflict between levels of selection'' which implies that a strategy may be optimal at the population level but not evolutionarily stable. To address this question, we may extend our model to treat strategies as variables that are themselves subject to evolution (Exercise~\ref{ex:evo}). For the model discussed in this chapter, the results show that strategies that optimize the long-term growth rate are indeed evolutionarily stable
(but this is no longer necessarily the case when considering, for instance, sexually reproducing populations~\cite{ZadorinRivoire2022}).

The same extension of the model to evolving strategies shows that knowledge of the statistics of the environment ($p_d$ for our example) is not required a priori but can effectively be learned through an evolutionary process. This solves a problem that appears also in gambling and finance where the statistics of the environment must be inferred from past experience. The question has been particularly studied in finance, where optimal learning strategies known as universal portfolios have been proposed~\cite{Cover1991}. In the simpler case of Kelly's model, the gambler may for instance record previous race results and use them together with Bayesian inference to predict the probability of the race outcomes \cite{DesponsLacoste2022}. With biological populations, however, learning must be performed at the individual level. One theoretical proposal that goes beyond random mutations is for instance that biological populations may use a reinforcement mechanism akin to Hebb's rule in neural learning~\cite{XueLeibler2016}. 


Finally, we note that the models that we presented rely on a strongly simplifying assumption: the environmental changes occur independently of the population. In fact, the environment is often also changing as the population grows, for instance through the consumption of nutrients. Even more generally, the environment may comprise other individuals from the same or other populations with which they may interact. This ecological dimension is the subject of other chapters.

\section*{Recommended readings}

\subsection*{Persistence} 
This reference describes the phenomenon of bacterial persistence.

Nathalie Q Balaban, Jack Merrin, Remy Chait, Lukasz Kowalik, and Stanislas Leibler.
Bacterial persistence as a phenotypic switch. Science, 305(5690):1622–1625, 2004. doi:
10.1126/science.1099390

\subsection*{Information theory for decision making under uncertainty} 

This book is a classic text on the use of information theory in problems from finance.

Thomas M. Cover and Joy A. Thomas. Elements of Information Theory. John Wiley \& Sons,
Hoboken, 2005. doi: 10.1002/047174882X

\subsection*{Model for information sensing}

The Kussel-Leibler model is one of the first to incorporate information theory into cellular behavior and signaling.

Edo Kussell and Stanislas Leibler. Ecology: Phenotypic diversity, population growth, and
information in fluctuating environments. Science, 309(5743):2075–2078, 2005. ISSN 00368075.
doi: 10.1126/science.1114383.

\section*{Exercises}

\begin{exercise}
\textbf{Kelly strategy with partial information}\label{ex:partialinfo1}
In analogy with Kelly's problem of betting on horse races, assume that different environments $E$ occur with independent probabilities $p(E)$ at each generation with a single phenotype $R=E$ permitting growth by a factor $f(E)$. In absence of any information, the optimal strategy $u(E)$ for long-term growth is proportional betting, $u(E)=p(E)$ (Box 2). Now assume that an information $S$ is available to each member of the population that relates to $E$ through a transition probability $q(S|E)$, i.e., $q(S|E)$ is the probability of perceiving $S$ given $E$. 

(1) Show that the long-term growth rate can be written in the form
\begin{equation}
\Lambda=\sum_S p(S)\left[\sum_E p(E|S)\ln (f(E)u(E|S))\right]
\end{equation}
where $p(S)$ is the probability to perceive $S$ averaged across all environments and $p(E|S)$ is the probability that environment is $E$ given that $S$ is perceived. Write $p(E|S)$ as a function of $p(E)$ and $q(S|E)$.

(2) Justify that the optimal strategy is $u(E|S)=p(E|S)$.

(3) Compare the optimal long-term growth rate in presence of information to the optimal growth rate in absence of information and show that the difference is given by the mutual information
\begin{equation}
I(E,S)=\sum_{E,S}q(S|E)p(E)\ln \frac{q(S|E)}{p(S)}
\end{equation}
The mutual information $I(E,S)$ therefore quantifies the value of information $S$ in this particular context.

(4) Acquiring information is generally costly. If the presence of the information channel $q(S|E)$ reduces the long-term growth rate by $c$, what are the conditions on $p(E)$ for the presence of this channel to be beneficial?

(5) The cost $c$ may be expected to depend on the precision of the sensor. Consider for instance a channel that reveals the correct environment with probability $1-\epsilon$ and otherwise does not reveal anything (so-called erasure channel). Given a cost $c(\epsilon)$ that increases when $\epsilon$ decreases, which value of $\epsilon$ provides an optimal trade-off between the value and the cost of information?
\end{exercise}

\begin{exercise}
\textbf{Value of information beyond Kelly's model}\label{ex:partialinfo2}
Consider the model of bacterial persistence introduced in the main text where cells can adopt two phenotypes, one growing irrespectively of the environment and the other growing only in absence of antibiotics.

(1) Express the long-term growth rate $\L$ in presence of an information $S$ modeled by an information channel $q(S|E)$.

(2) What is the optimal strategy given $S$?

(3) Show by comparing to a situation with no information that the value of information can be strictly lower than $I(S,E)$.
\end{exercise}

\begin{exercise}
\textbf{Stochastic sensing at the level of individual cells}\label{ex:partialinfo3}
In the two previous exercises, the information $S$ is assumed to be common to each member of the population. Here we assume instead that each individual has its own sensor $q(S|E)$ so that $S$ may differ from one individual to the next.

(1) Justify that in this case the long-term growth rate takes the form
\begin{equation}
\Lambda=\sum_E p(E)\ln \left(\sum_{R,S} f(R,E)u(R|S)q(S|E))\right)
\end{equation}
(2) Use the concavity of the logarithm (Jensen's inequality) to justify that the same information channel $q(S|E)$ has more value at the individual level than at the population level. 
\end{exercise}

\begin{exercise}
\textbf{Pareto front for Kelly's model}\label{ex:pareto}
Let us consider Kelly's model with fair odds for two horses. Let the probability that the first horse wins be $p$, the bet and the odd on the first horse be $b$ and $1/r$. 

(1) Write the expression of the mean growth rate $\langle W \rangle$, and of the volatility  $\sigma_W$ for this problem. 
Show that there is a risk free strategy when $b=r$.

One introduces the objective function 
\begin{equation}
J= \alpha \langle W \rangle  - (1-\alpha) \sigma_W.
\end{equation}

(2) From the optimization of $J$ show that the optimal strategy has the two branches 
shown in Fig.~\ref{fig:2horses}. 
Show that the optimal bets on these two branches are of the form $b^\pm = p \pm \gamma \sigma$, 
where $\gamma=(1-\alpha)/\alpha$ and $\sigma=\sqrt{p(1-p)}$.

(3) Show that the slope of the Pareto border has the form
\begin{equation}
\frac{d \sigma_W}{d \langle W \rangle}= \frac{\sigma}{p-b}.
\end{equation}
What happens to this slope near Kelly's point and near the risk free strategy ?
\end{exercise}

\begin{exercise}
\textbf{Evolution of an optimal strategy}\label{ex:evo}
Here we consider evolving the strategy itself.

(1) Implement numerically the model of bacterial persistence introduced in the main text for a large but finite population. To this end, consider $N$ individuals (e.g., $N=1000$), each with an attribute $R$. For each individual, draw a random number $\xi$ of descendants, with mean $f(R,E_t)$ where $E_t$ drawn from $P(E)$ is common to all individuals. Assign a $R$ to each of these descendants with probability $u(R)$. If the total number of descendants $N_t$ is non-zero, record the ratio $N_t/N$ and re-sample at random the population to bring back its size to $N$. Show that provided that $N$ is large enough and $N_t$ does not reach $0$ then $(\sum_t\ln(N_t/N))/t$ provides a good approximation to the growth rate $\Lambda$ in the limit of large $t$.

(2) Extend the model to make $u_du(R={\rm dormant})$ an attribute of each individual. Assume that $u_d$ is transmitted from one parent to one of its offspring as $u_d=\min(1,\max(0,u_d+\mu))$ where $\mu$ is normally distributed with variance $\s_M^2$. Show that provided that $\s_M^2$ is small enough, the distribution of $u_d$ evolves to be centered around the optimal $u_d$. 
\end{exercise}

\hideinbooklayout{
\ \\
\noindent\href{https://creativecommons.org/licenses/by-sa/4.0/}{\includegraphics[width=2.5cm]{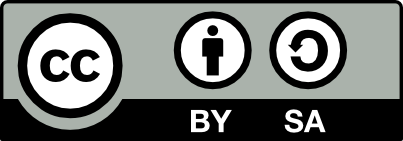}}
}

\ifx\weAreInMain\undefined                                                      %
\bibliographystyle{unsrtnat}                                                    %
\bibliography{\chapterroot bibliography-chapter}                                %
\end{document}                                                                  %
\fi                                                                             %

%% file: main/macros.tex
\usepackage[utf8]{inputenc}
\usepackage[T1]{fontenc}
\usepackage{amsmath}
\usepackage{amssymb}
\usepackage{epsfig}
\usepackage{float}
\usepackage{geometry}
\usepackage{graphicx}
\usepackage{listings}
\usepackage{lmodern}
\usepackage{mdframed}
\usepackage[numbers]{natbib}
\usepackage{siunitx}%
\usepackage{xcolor}
\usepackage{pgfplots}
\usepackage{mathtools}
\usepackage{physics}
\usepackage{setspace}
\usepackage{chngcntr}
\usepackage{enumitem}

\usepackage[nottoc,notlot,notlof]{tocbibind} 

\usepackage[textsize=footnotesize,textwidth=2.8cm]{todonotes}
\usepackage[version=4]{mhchem}
\ifx\tmpmacrosfolder\undefined
  \newcommand{\tmpmacrosfolder}{.}%
\fi
\ifx\weAreInMain\undefined
{}
\else
\input{\tmpmacrosfolder/tmp_macros_chapters.tex}%
\fi
\usepackage[hidelinks]{hyperref}
\definecolor{darkblue}{rgb}{0.15,0.,0.65}
\hypersetup{
	colorlinks=true,
	linkcolor=darkblue,
	filecolor=magenta,
        citecolor=darkblue,
	urlcolor=darkblue,
	pdftitle={Economic Principles in Cell Biology},
}
\newcommand{\includechapterauthors}[1]{%
\par\noindent%
#1\\
}%

\newcommand{\chapterroot}{./chapters/TEMPLATE/}

\newcommand{\copyrightinfo}[1]
  {{\color{magenta}\textsc{Copyright Info:} #1}}

\newcommand{\hideinbooklayout}[1]{#1}

\floatstyle{plain}%
\newfloat{EPCPFloatBlock}{b}{loepcpb}[chapter]

\mdfdefinestyle{EPCPBlocks}{%
  linecolor=blue!50!black,%
  frametitlerule=true,%
  frametitlebackgroundcolor=\EPCPBlockTitleColor,%
  backgroundcolor=\EPCPBlockBGColor,%
  innertopmargin=9pt,
  innerbottommargin=9pt,
  skipabove=7pt,
  skipbelow=7pt
}
\newmdenv[style=EPCPBlocks]{EPCPBlockMD}%

\makeatletter
\newcommand{\EPCPBlockTitleFont}{\normalfont\bfseries}%
\definecolor{EPCPviolet}{rgb}{0.15,0.,0.65}
\newcommand{\EPCPBlockTitleFontColor}{EPCPviolet}
\newcommand{\EPCPBlockTitleColor}{EPCPviolet!10}
\newcommand{\EPCPBlockBGColor}{EPCPviolet!3}
\newmdenv[style=EPCPBlocks,frametitle={\color{\EPCPBlockTitleFontColor}Chapter highlights}%
]{chapterhighlights}

\newcommand{\EPCP@FloatSpec}{X}%
\newcommand{\EPCP@BlockTitle}{}%
\define@key{EPCPBlocks}{label}{\label{#1}}%
\define@key{EPCPBlocks}{title}{\renewcommand{\EPCP@BlockTitle}{:~{\EPCPBlockTitleFont#1}}}%
\define@key{EPCPBlocks}{float}{\renewcommand{\EPCP@FloatSpec}{#1}}%
\newcounter{ecoanaboxcounter}
\counterwithin{ecoanaboxcounter}{chapter}
\renewcommand{\theecoanaboxcounter}{\arabic{chapter}.\Alph{ecoanaboxcounter}}
\newenvironment{EPCPBlock}[3][]%
{
\setkeys{EPCPBlocks}{#1}%
\ifthenelse{\equal{\EPCP@FloatSpec}{X}}
{
\begin{EPCPBlockMD}%
[frametitle={\color{\EPCPBlockTitleFontColor}#2#3\EPCP@BlockTitle}]
}
{
\floatplacement{EPCPFloatBlock}{\EPCP@FloatSpec}%
\begin{EPCPFloatBlock}\begin{EPCPBlockMD}%
  [frametitle={\color{\EPCPBlockTitleFontColor}#2#3\EPCP@BlockTitle}]%
}%
}{
\ifthenelse{\equal{\EPCP@FloatSpec}{X}}
{
\end{EPCPBlockMD}
}
{
\end{EPCPBlockMD}\end{EPCPFloatBlock}
}
}

{\begin{EPCPBlock}[#1]{Economic analogy\mbox{~}}{\refstepcounter{ecoanaboxcounter}\theecoanaboxcounter}}
{\end{EPCPBlock}}
\newenvironment{Ecoblock*}[1][]%
{\begin{EPCPBlock}[#1]{Economic analogy}{}}
{\end{EPCPBlock}}
\newcounter{philquestboxcounter}
\counterwithin{philquestboxcounter}{chapter}
\renewcommand{\thephilquestboxcounter}{\arabic{chapter}.\Alph{philquestboxcounter}}
{\begin{EPCPBlock}[#1]{Philosophical remarks\mbox{~}}{\refstepcounter{philquestboxcounter}\thephilquestboxcounter}}
{\end{EPCPBlock}}
\newenvironment{Philblock*}[1][]%
{\begin{EPCPBlock}[#1]{Philosophical remarks}{}}
{\end{EPCPBlock}}
\newcounter{expmethboxcounter}
\counterwithin{expmethboxcounter}{chapter}
\renewcommand{\theexpmethboxcounter}{\arabic{chapter}.\Alph{expmethboxcounter}}
{\begin{EPCPBlock}[#1]{Experimental methods\mbox{~}}{\refstepcounter{expmethboxcounter}\theexpmethboxcounter}}
{\end{EPCPBlock}}
\newenvironment{Expblock*}[1][]%
{\begin{EPCPBlock}[#1]{Experimental methods}{}}
{\end{EPCPBlock}}
\newcounter{mathdetailboxcounter}
\counterwithin{mathdetailboxcounter}{chapter}
\renewcommand{\themathdetailboxcounter}{\arabic{chapter}.\Alph{mathdetailboxcounter}}%
\newenvironment{MathDetailblock}[1][]%
{\begin{EPCPBlock}[#1]{Mathematical details\mbox{~}}{\refstepcounter{mathdetailboxcounter}\themathdetailboxcounter}}
{\end{EPCPBlock}}
\newenvironment{MathDetailblock*}[1][]%
{\begin{EPCPBlock}[#1]{Mathematical details}{}}
{\end{EPCPBlock}}

\newcommand{\EPCP@GeneralBlockTitle}{A}%
\define@key{EPCPGeneralblock}{title}{\renewcommand{\EPCP@GeneralBlockTitle}{{\EPCPBlockTitleFont#1}}}%

\makeatother

\newcounter{exercisecntr}
\counterwithin{exercisecntr}{chapter}
\newenvironment{exercise}
  {%
   \par\noindent%
   \refstepcounter{exercisecntr}%
   \textbf{Exercise \theexercisecntr}}
  {\mbox{}}

%% file: macros-chapter.tex
\newcommand{\g}{\gamma}
\newcommand{\s}{\sigma}
\renewcommand{\L}{\Lambda}
\newcommand{\dkl}{D_{\text{KL}}}

\newcommand{\E}{\mathbb{E}}
\newcommand{\lp}{\log\!\text{-}\mathrm{cap}}
\newcommand{\p}{\mathrm{p}}
\newcommand{\q}{\mathrm{q}}
\newcommand{\R}{\mathrm{r}}
\newcommand{\B}{\mathrm{b}}
\newcommand{\odds}{\mathrm{o}}

\renewcommand{\underline}[1]{\mathbf{#1}}

\newcommand{\be}{\begin{equation}}
\newcommand{\ee}{\end{equation}}

%% file: style-chapter.tex
\usepackage{setspace}
\usepackage[final
           ]{showkeys}
\definecolor{refkey}{rgb}{.65,.65,.65}
\definecolor{labelkey}{rgb}{.65,.65,.65}

\usepackage{enumitem}
\setlist{noitemsep}